\documentclass[sort&compress,
  ,draft            
  ]
  {aipproc}

\layoutstyle{8x11single}
\usepackage{xspace}
\usepackage{amsmath}
\usepackage{epstopdf}
\usepackage{graphicx}
\usepackage{subfig}
\usepackage{caption}
\pdfoutput=1

\newcommand{\invfb}{\ensuremath{\rm fb^{-1}}\,}
\newcommand{\simle}{\mathrel{
   \rlap{\raise 0.511ex \hbox{$<$}}{\lower 0.511ex \hbox{$\sim$}}}}
\newcommand{\mett}{\mbox{$E\!\!\!\!/_{T}$}\xspace}
\newcommand {\asld} {\ensuremath{a^d_{\mathrm{sl}}}\xspace}
\newcommand {\asls} {\ensuremath{a^s_{\mathrm{sl}}}\xspace}
\newcommand{\gevcc}{\ensuremath{\mathrm{GeV}/c^2}\xspace}
\newcommand{\tevcc}{\ensuremath{\mathrm{TeV}/c^2}\xspace}
\begin{document}

\title{Searches for New Physics at the Tevatron and LHC}

\classification{12.60.-i, 13.85.Rm, 14.65.Jk, 14.80.-j}

\keywords{Tevatron, LHC, particle physics, proton-proton
  collisions, proton-antiproton collisions, collider physics} 

\author{Peter Wittich$^a$,
  on behalf of the ATLAS, CDF, CMS and D0 Collaborations}{
  address={$^a$Department of Physics, Cornell University,
    Ithaca, NY 14853, USA,
    {\tt wittich@cornell.edu}}
}

\begin{abstract}
  This is an auspicious moment in experimental particle physics -
  there are large data samples at the Tevatron and a new energy regime
  being explored at the Large Hadron Collider with ever larger data
  samples. The coincidence of these two events suggests that we will
  soon be able to address the question, what lies beyond the standard
  model? Particle physics's current understanding of the universe is
  embodied in it. The model has been tested to extreme precision -
  better than a part in ten thousand - but we suspect that it is only
  an approximation, and that physics beyond this standard model will
  appear in the data of the Tevatron and LHC in the near future.  This
  brief review touches on the status of searches for new physics at
  the time of the conference.
\end{abstract}

\maketitle

The PANIC conference occurred during an exciting time in particle
physics. The Tevatron experiments were evaluating results with up to
9~\invfb of data, an amount inconceivable in years past.  The LHC had
collected its first 1~\invfb of data and results were being presented
on datasets 30 times larger than those published in the previous year.
As of the time of the conference no evidence for new physics had been
seen, but the prospect for discovery is clear - in the meantime, the
LHC has accumulated more than five times the data set than was
presented at the conference, and next year's data set promises to be
bigger by another factor of 2-4.


Due to the large number of results that are available, the plenary
talk only presented a small number of the results that were available
at the time. Many others were shown in the parallel sessions. In
addition to the results discussed below, the rest of the results are
available on the web~\cite{atlas_public_results,cms_public_webpage,
  cdf_public_results,d0_public_results}.

This article is organized as follows. After a brief introduction to
the experimental facilities, selected results will be recapped with
references to more information. We encourage the reader to go to these
sources for more information. Due to space constraints very few
details can be included in this paper.

\subsection{Experimental Facilities: Tevatron at FNAL}
The Tevatron is a proton-antiproton collider at the Fermi National
Accelerator Laboratory (FNAL), located outside Chicago, Illinois in
the US~\cite{tevatron}. The Tevatron has been taking data for over 25
years and 2011 was the last year of the proton-antiproton program at
Fermilab.  During the Run 2 data-taking period, data equivalent to
12~\invfb were delivered to the experiments.
The center-of-mass energy of the collisions in Run 2 is 1.96~TeV. The
Tevatron has two active interaction points, where general-purpose
collider detectors CDF and D0 are housed.

\paragraph{CDF} CDF is a multi-purpose experiment with an emphasis on
charged-particle tracking.  The CDF~II detector is described in detail
elsewhere~\cite{CDF_detect_A}.  It has a solenoidal charged particle
spectrometer, consisting of 7-8 layers of silicon microstrip detectors
and a cylindrical drift chamber immersed in a 1.4~T solenoidal
magnetic field, a segmented sampling calorimeter, and a set of charged
particle detectors outside the calorimeter used to identify muon
candidates.

\paragraph{D0} The Run 2 D0 detector consists of three major
components: a central tracker, liquid argon/uranium calorimeters, and
a muon spectrometer. The tracking system consists of a silicon strip
tracker and a fiber tracker. Both are enclosed in a 2~T solenoidal
magnetic field.  The muon system consists of gaseous detectors with
toroidal magnetic fields to assist in the momentum measurement. More
details can be found elsewhere~\cite{d0tdr}.

\subsection{Experimental Facilities: CERN LHC}
The Large Hadron Collider (LHC) is a proton-proton collider at the
CERN Laboratory in Geneva, Switzerland~\cite{lhcmachine}.  There are
four collision regions at the LHC. Two of the four are occupied by the
general-purpose experiments, CMS and ATLAS. The other two are occupied
by LHCb, an experiment dedicated to studying the production and decays
of $B$ hadrons and mesons, and ALICE, an experiment dedicated to
heavy-ion collisions.  Compared to the Tevatron experiments, ATLAS and
CMS are next-generation detectors. Their fiducial coverage extends
further in $\eta$, and most systems have higher segmentation to handle
the more demanding environment of the LHC.  At design performance, the
LHC will collide bunches every 25~ns (compared to 396~ns for the
Tevatron) at an energy of $\sqrt{s}=14$~TeV (compared to just under
2~TeV at Fermilab.)  The higher bunch intensities will lead to more
than five times higher pileup, i.e., five times higher number of extra
interactions per crossing, as well, and the overall design
instantaneous luminosity of $10^{34}/\rm{cm}^2/\rm{s}$ is also 25
times higher than the maximum achieved at Fermilab.

\begin{figure}[tbp]
  \centering
  \includegraphics[width=0.4\textwidth]{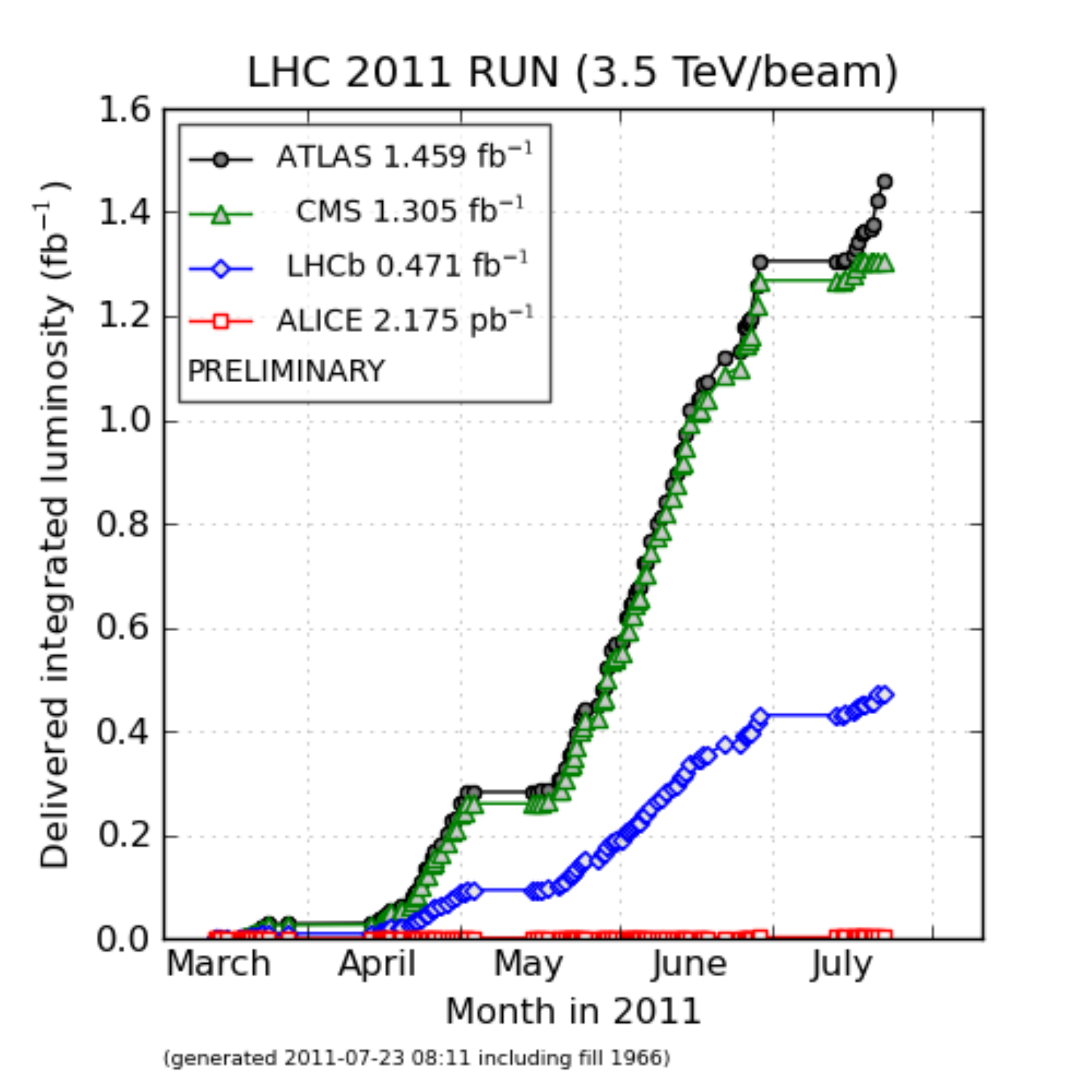}
  \includegraphics[width=0.4\textwidth]{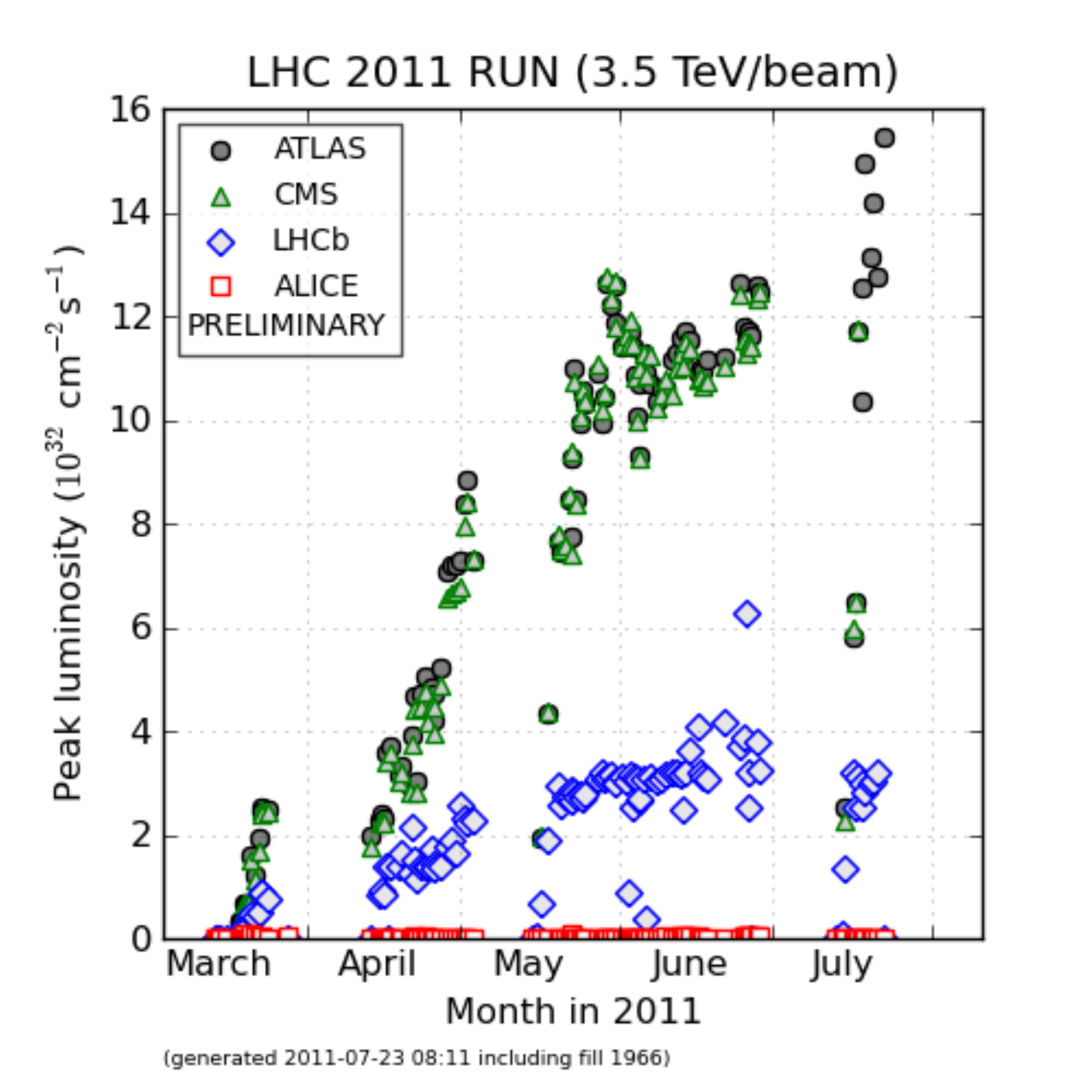}
  \caption{LHC luminosity as a function of calendar day at the time of
  the PANIC11 conference. Different colors represent different
  interaction regions at the LHC. The luminosity at ALICE and LHCb is
  not optimized for the overall largest luminosity.}
  \label{fig:lhclumi}
\end{figure}
The current data-taking period started in December 2009. In the
calendar year 2010, 35/pb of data were collected at a center-of-mass
energy of 7~TeV; at the time of PANIC11, more than 1~\invfb had been
collected in 2011 at the same energy. Figure~\ref{fig:lhclumi} shows
the accumulation of data as a function of time as well as the highest
recorded instantaneous luminosity at the time of the conference
(roughly $1.6\times10^{33}/\rm{cm^2}/s$). More than 45/pb of data were
being accumulated \emph{per day}. In the meantime, instantaneous
luminosities as high as $3.7\times10^{33}/\rm{cm^2}/s$ have been
recorded, and more than 130/pb of data have been accumulated in a
single store (which is roughly equivalent to a day).  For the calendar
year 2011, the LHC goal was to deliver 1~\invfb of data. This goal was
already achieved by the time of the conference, and more than 5~\invfb
of data have been delivered in 2011.


\paragraph{CMS}
The Compact Muon Solenoid (CMS) detector is one of two large
general-purpose experiments at the LHC.  The 12,500 ton detector
features nearly complete solid-angle coverage and can precisely
measure electrons, photons, muons, jets and missing energy over a
large range of particle energies.  A detailed description of the CMS
detector can be found elsewhere~\cite{cmsjinst:2008}.  The detector
consists of the pixel and silicon strip trackers. They provide
coverage in the region $|\eta|<2.5$ and are immersed in a $3.8$~T
magnetic field to allow momentum determination of charged
particles. The electromagnetic and hadron calorimeters are used to
detect energy deposits from electrons in the range $|\eta|<2.5$ as
well as to provide an estimate of missing transverse energy due to
escaping particles. The electromagnetic calorimeter has an energy
resolution of better than 0.5\,\% for unconverted photons with
transverse energies above 100~GeV. 
A set of gaseous detectors outside the solenoids is used to detect the
passage of minimum-ionizing particles, mainly muons.

\paragraph{ATLAS} The other large general-purpose detector is
ATLAS. ATLAS is described in detail elsewhere~\cite{atlasjinst}.  The
tracking system consists of a silicon pixel detector, a silicon strip
tracker, a straw chamber with transition radiation particle ID. The
tracker is bathed in a 2~T solenoidal magnetic field. Outside of the
magnetic field, electromagnetic and hadronic calorimeters allow
measurement of the energy deposited by passing particles.  There are
three toriodal magnetic fields (two in the endcaps, one in the central
region) that produce magnetic fields for the muon systems of 1~T and
0.5~T, respectively. The muon system consists of separate
high-precision tracking chambers and triggering system, with the goal
of achieving 10\% resolution for 1~TeV muons.

\section{Searches for New Physics}

The program of searches for new physics at collider experiments is
exhaustive. We look for new physics either in direct or in indirect
searches. In direct searches, the goal is to look for evidence of
on-shell production of massive undiscovered particles. These particles
can be strongly produced and can decay either strongly, to hadronic
final states, or weakly, to leptonic final states. A more complicated
possibility includes some strong and some weak decays in complicated
decays when multiple new particles are produced, leading to mixed
hadronic and leptonic final states.

In indirect searches, rather than look for the direct production of a
heavy new particle, we make precision measurements of quantities that
can be affected by the presence of new particles, e.g. through
interference. These effects often come through loop diagrams, where
one internal line in a loop is replaced by a new particle. The
presence of the new particle changes the measured quantity, such as a
branching ratio, from its standard model expectation.  Since the new
particles that are produced are typically off-shell, these searches
are able to probe much higher masses than the direct
searches. However, they are dependent on the theoretical calculations
and good understanding of systematic uncertainties.

Direct searches can be characterized by the final state
particles. Many of the most sensitive early searches at the LHC
involve strongly produced particles (which have large cross sections)
that decay to hadronic final states (with large branching fractions.)
The challenge with these searches is to suppress the equally large
multi-jet background. Some of these searches are discussed below,
including narrow dijet resonances, jets + \mett SUSY searches, and
searches for SUSY in events with $b$ jets and \mett.

\paragraph{Narrow resonances in dijet events}
\begin{figure}[hbtp]
  \centering
  \includegraphics[width=0.33\textwidth]{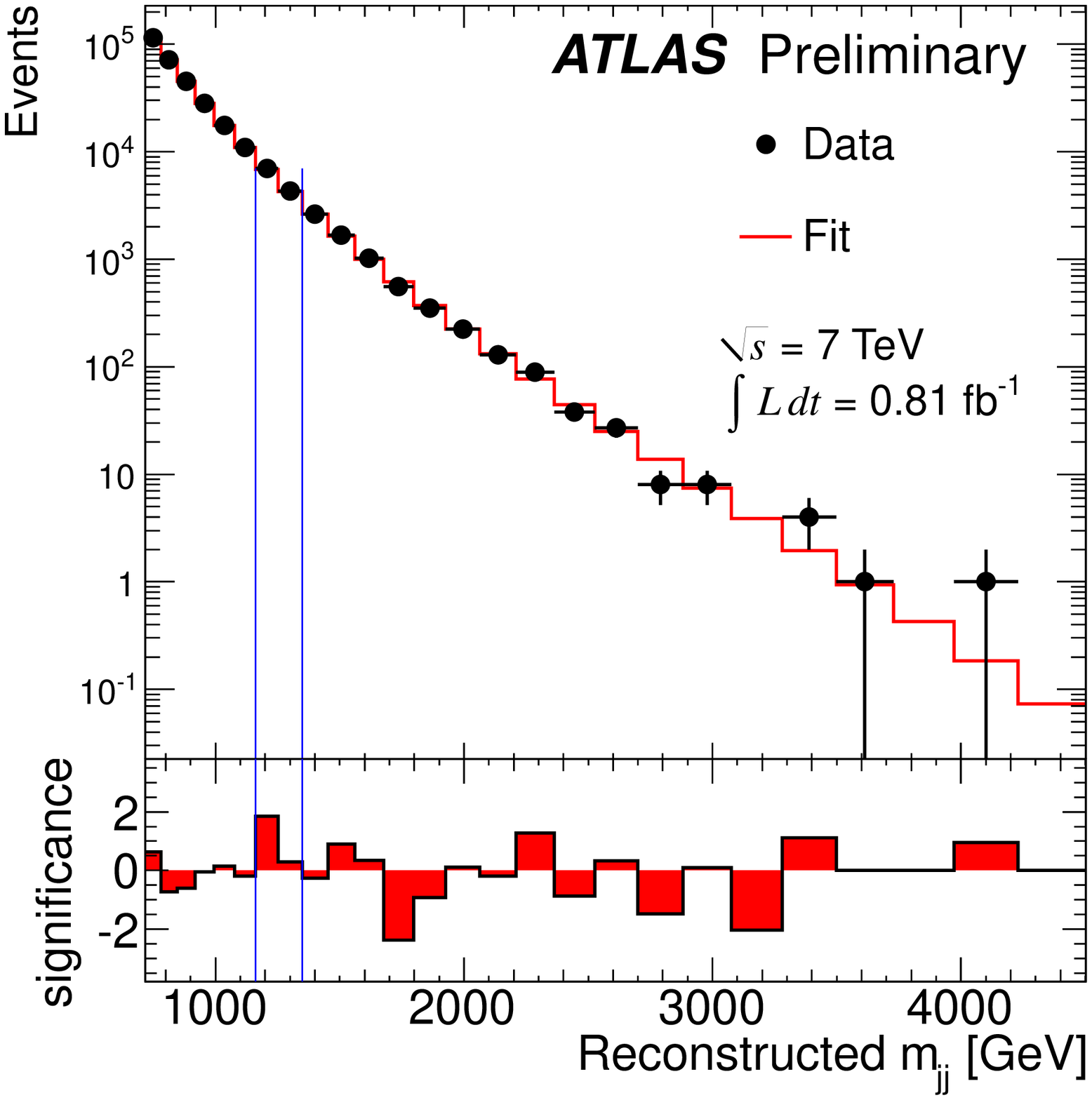}
  \includegraphics[width=0.33\textwidth]{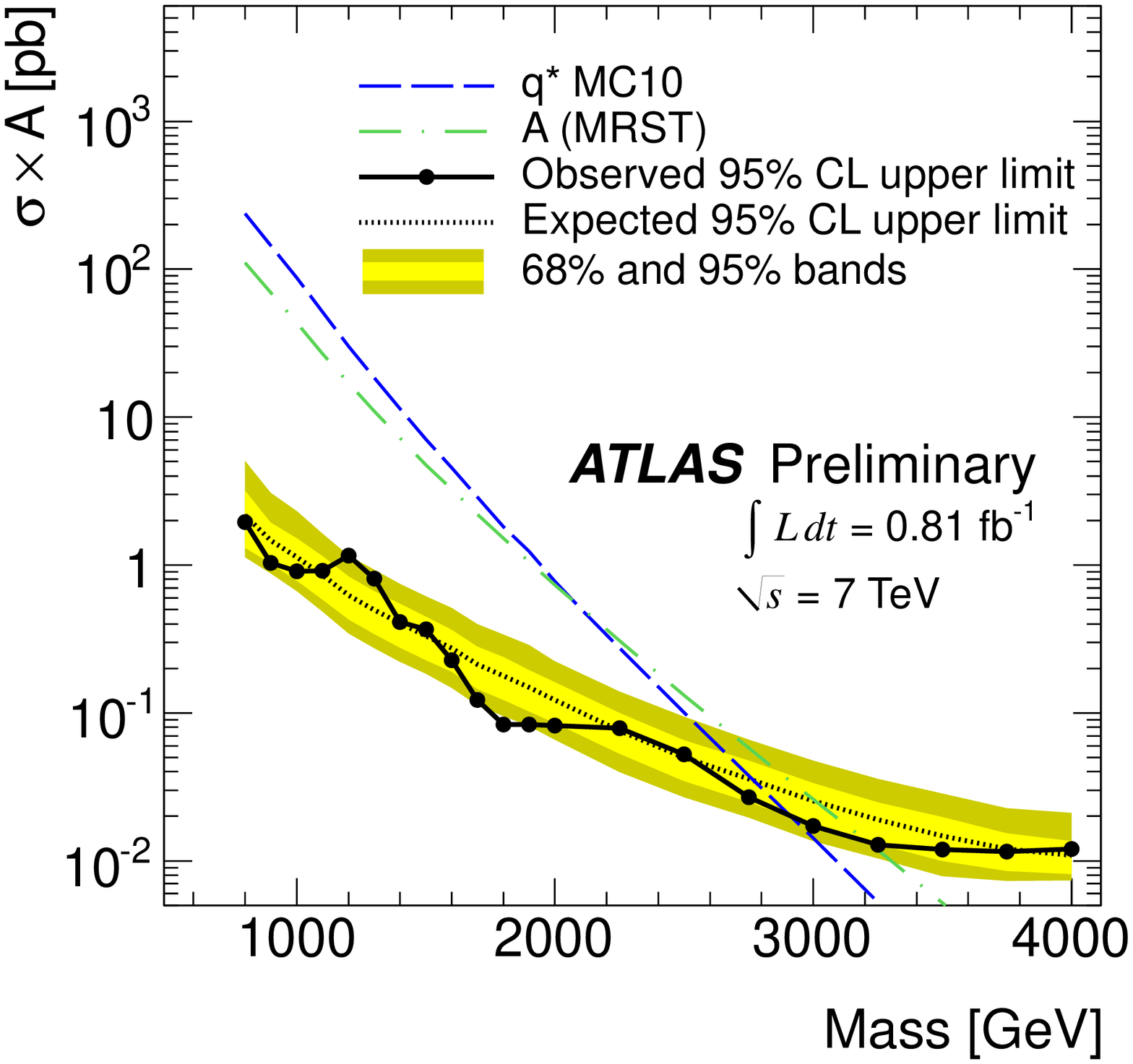}
  \includegraphics[width=0.33\textwidth]{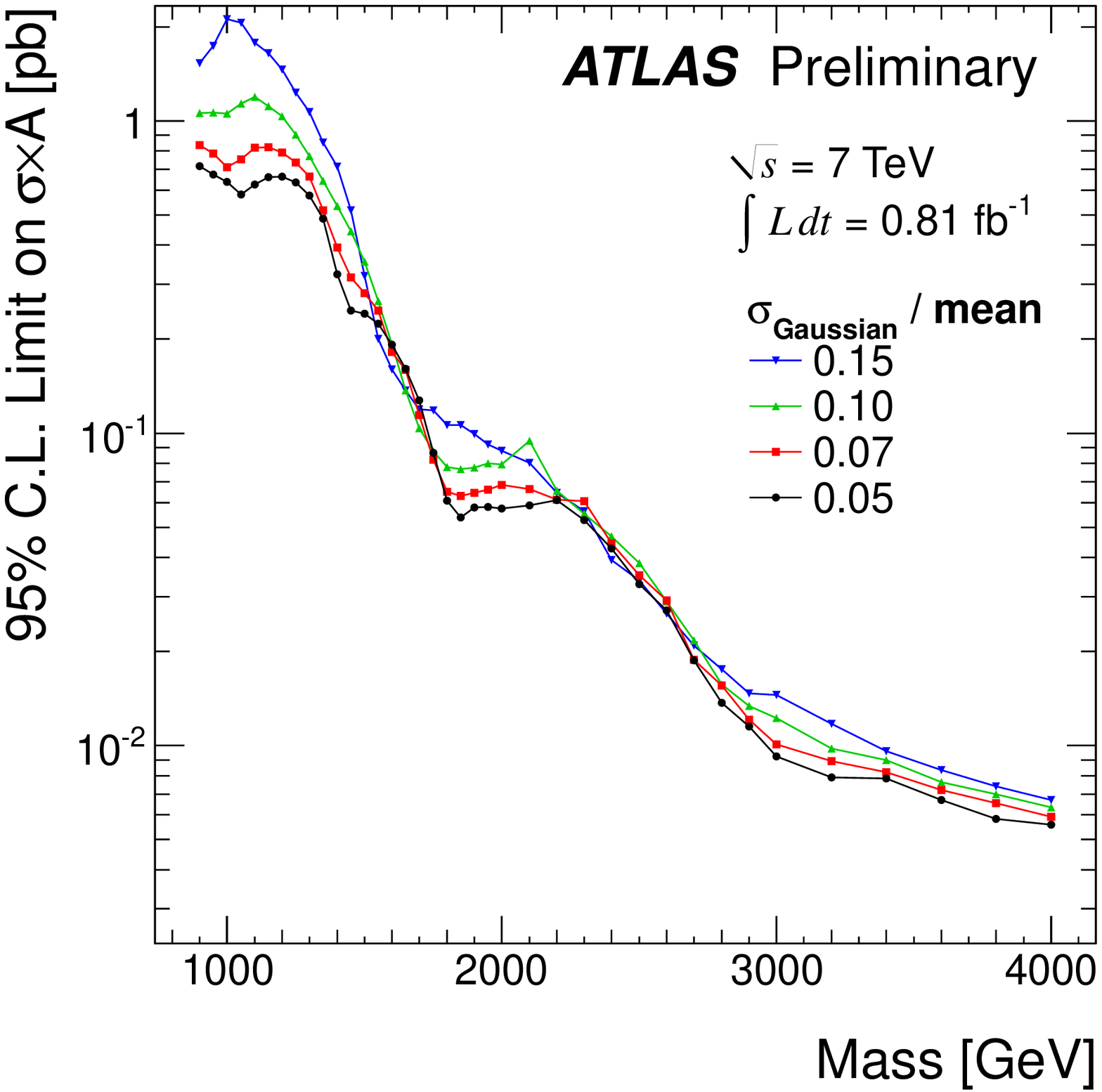}
  \caption{ATLAS results for searches for narrow dijet resonances in
    0.81~\invfb of data. Left: the invariant mass spectrum of pairs of
    jets selected by this analysis. No significant excess is
    observed. Middle: limit curves for $\sigma\times A$ for excited
    quark models. Right: limit curves for $\sigma\times A$ for simple
    Gaussian resonances decaying to dijets as a function of mass, for
    values of the fractional Gaussian width. }
  \label{fig:atlas_dijetbump}
\end{figure}
For the conference, ATLAS presented a new search for resonant dijet
production of new particles. Using 0.81~\invfb of data, the ATLAS
collaboration calculates the invariant mass of pairs of jets (see
Fig.~\ref{fig:atlas_dijetbump} left) and looks for structure in the
falling spectrum indicative of resonant $s$-channel production of new
particles (e.g., $gg\to X \to qq$)~\cite{ATLAS-CONF-2011-095}. These
searches are sensitive to many new physics models such as excited
quarks and contact interactions, and are also sensitive to quark
substructure, and are therefore the modern equivalent of Rutherford
scattering - relevant since the conference was dedicated to the
Rutherford centennial. No significant excess was observed, and the
collaboration presented the results as limits on the cross section
times branching ratio times acceptance for both excited quarks
(Fig.~\ref{fig:atlas_dijetbump} middle) and for simple parameterized
Gaussian shapes (Fig.~\ref{fig:atlas_dijetbump} right.) In the latter
case, the results are presented as a function of the mean of the
Gaussian for several fractional widths (5-15\%). Using these generic
assumptions, limits are placed in the range of $1 - 10^{-2}$~pb for
masses in the range of $1-3~\mathrm{TeV}/c^2$.  CMS has released similar
results~\cite{cmsdijets}.

\paragraph{Squarks and Gluinos}
In supersymmetric models, $\tilde{g}\tilde{g}$, $\tilde{g}\tilde{q}$
and $\tilde{q}\tilde{q}$ production leads to hadronic final states. If
$R$-parity is assumed to be conserved, the lightest supersymmetric
particle is stable and usually assumed to be the dark matter
candidate, \emph{i.e.}, neutral. In this case, the final state
contains missing transverse momentum (\mett)\footnote{ We define the
  missing transverse momentum $\vec{\mett}$$\equiv -\sum_i E_{\rm T}^i
  {\bf n}_i$, where ${\bf n}_i$ is the unit vector in the azimuthal
  plane that points from the beamline to the $i$th calorimeter
  tower. We call the magnitude of this vector \mett.  } due to the
escaping undetected particles.  As such the final state consists of
jets and large \mett. The analysis uses 1~\invfb of data collected
using a trigger requiring a jet with $E_T>75$~GeV and
$\mett>45$~GeV. The measurement is optimized for different squark and
gluino final states, and the final number of event counts is extracted
by a likelihood fit to the a mass-like variable ($m_\text{eff}$),
which is the sum of \mett and the magnitudes of the transverse momenta
of the highest momentum jets considered in the analysis. No excess
above SM expectation is observed, and limits are set in two ways: an
mSUGRA scenario, and a simplified model where the only free parameters
are the masses of the squarks and gluinos and the LSP mass is
assumed to be zero.
\begin{figure}[htbp]
  \centering
  \includegraphics[width=0.4\textwidth]{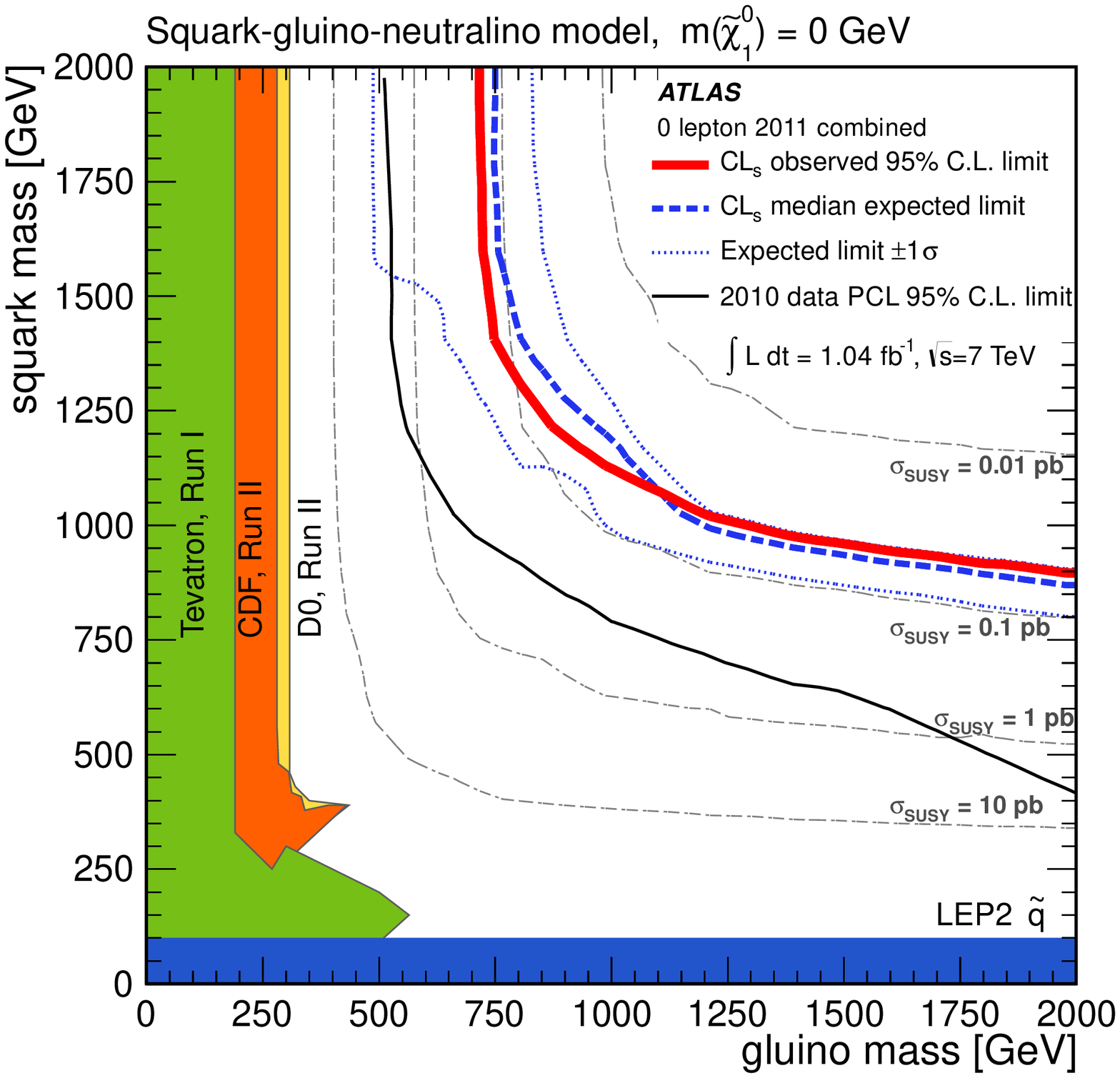}
  \includegraphics[width=0.4\textwidth]{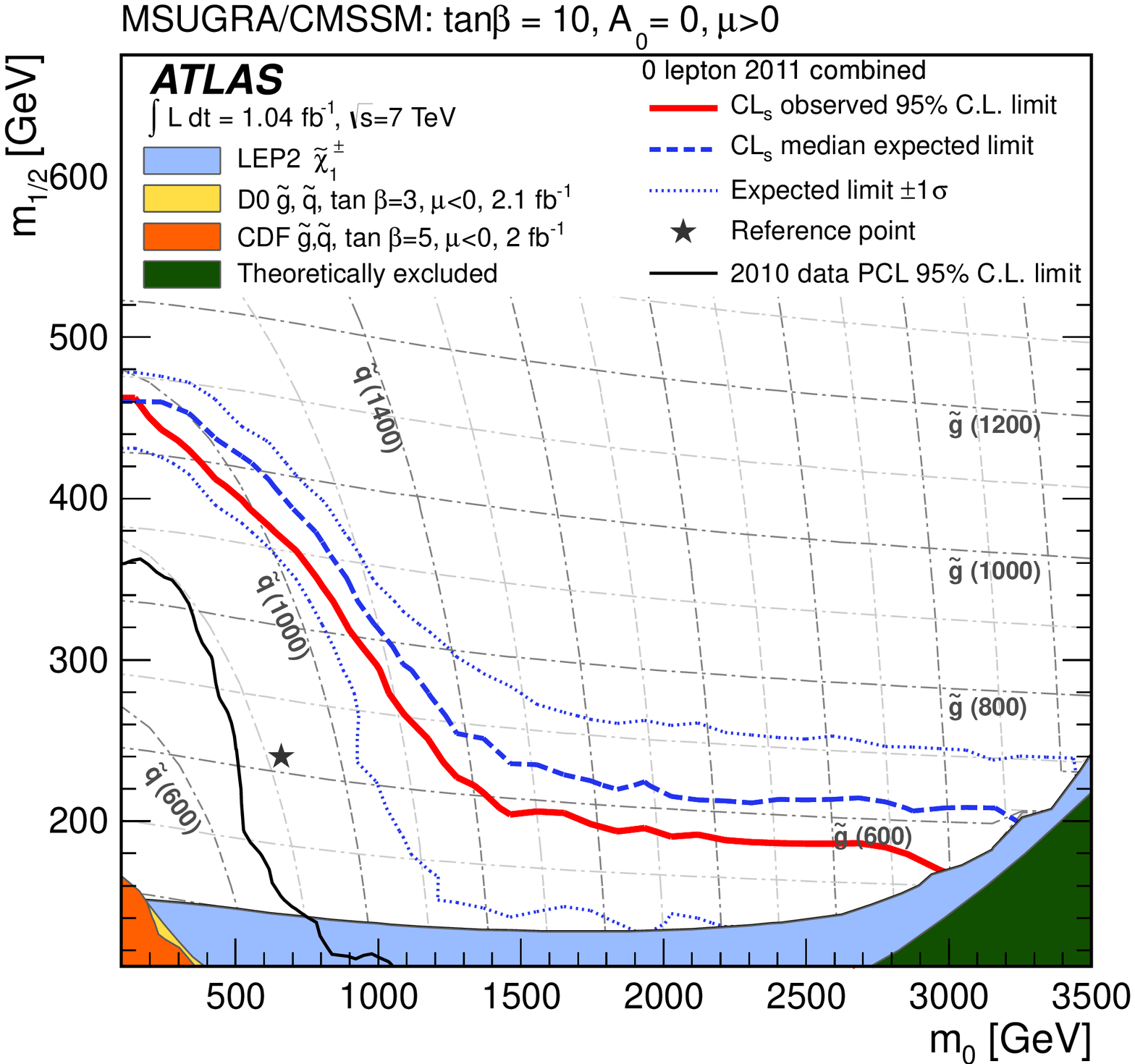}
  \caption{ATLAS: Interpretation of null results for searches for new
    physics with jets + \mett in a simplified model (left) and mSugra
    (right). In the simplified models, gluino (squark) masses below
    $700 (875) \gevcc$ are excluded at the 95\% confidence level,
    while in the mSugra-inspired model squarks and gluinos of equal
    mass are disallowed for masses below 950~\gevcc($\tan\beta=10$,
    $A_0=0$ and $\mu> 0$). }
  \label{fig:atlas_jets_met_contour}
\end{figure}
The results can be seen in Fig.~\ref{fig:atlas_jets_met_contour}.  In
the simplified models, gluino (squark) masses below 700 (875)\gevcc are
excluded at the 95\% confidence level, while in the mSUGRA-inspired
model squarks and gluinos of equal mass are disallowed for masses
below 950 \gevcc ($\tan\beta=10$, $A_0=0$ and $\mu>
0$)~\cite{atlas_2011_susy_jetsmet}.

\begin{figure}[htbp]
  \centering
  \begin{minipage}{0.5\linewidth}
    \vspace{0pt}
    \includegraphics[width=\textwidth]{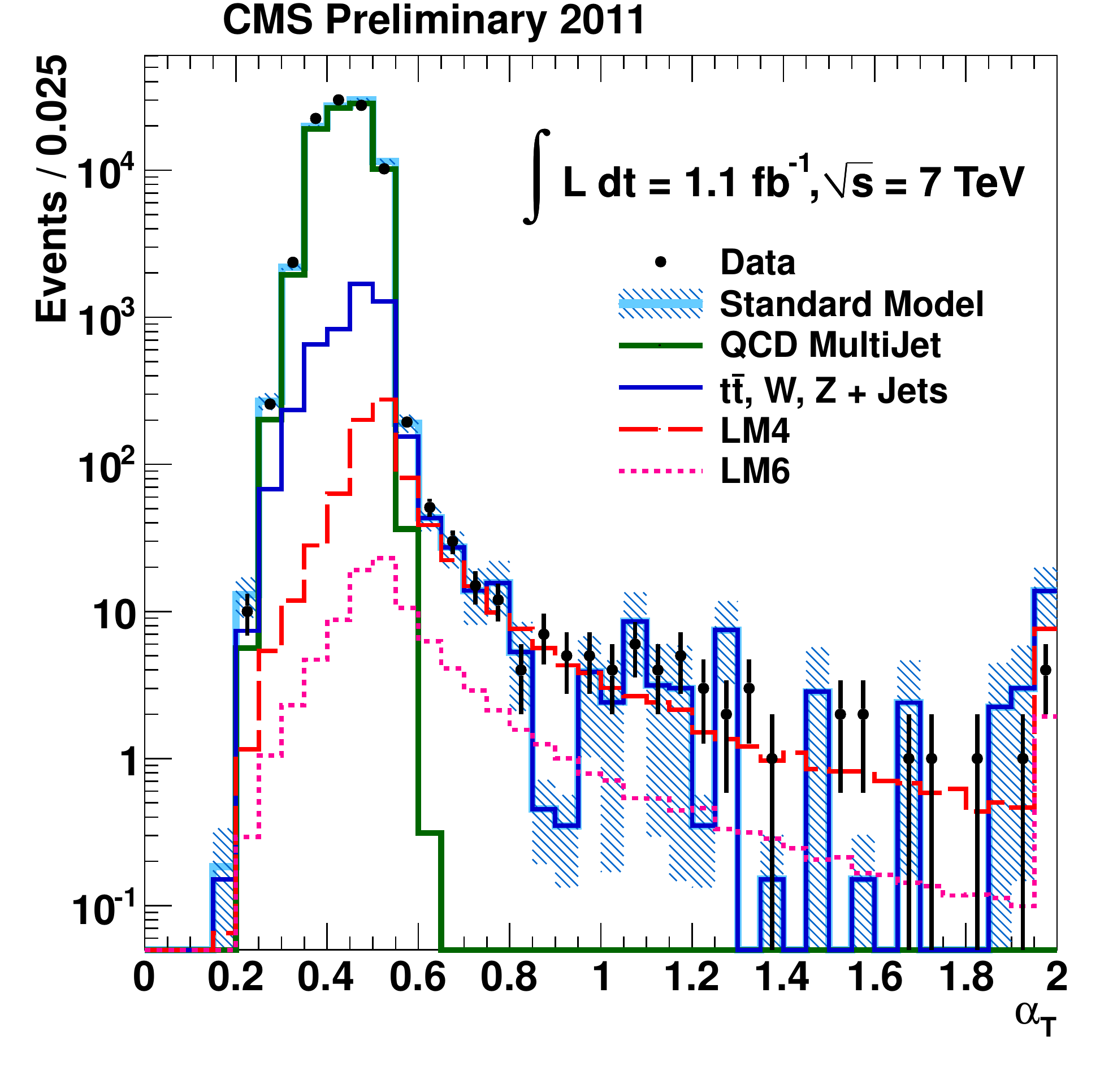}
  \end{minipage}
  \hfill
  \begin{minipage}{0.5\linewidth}
    \includegraphics[width=\textwidth]{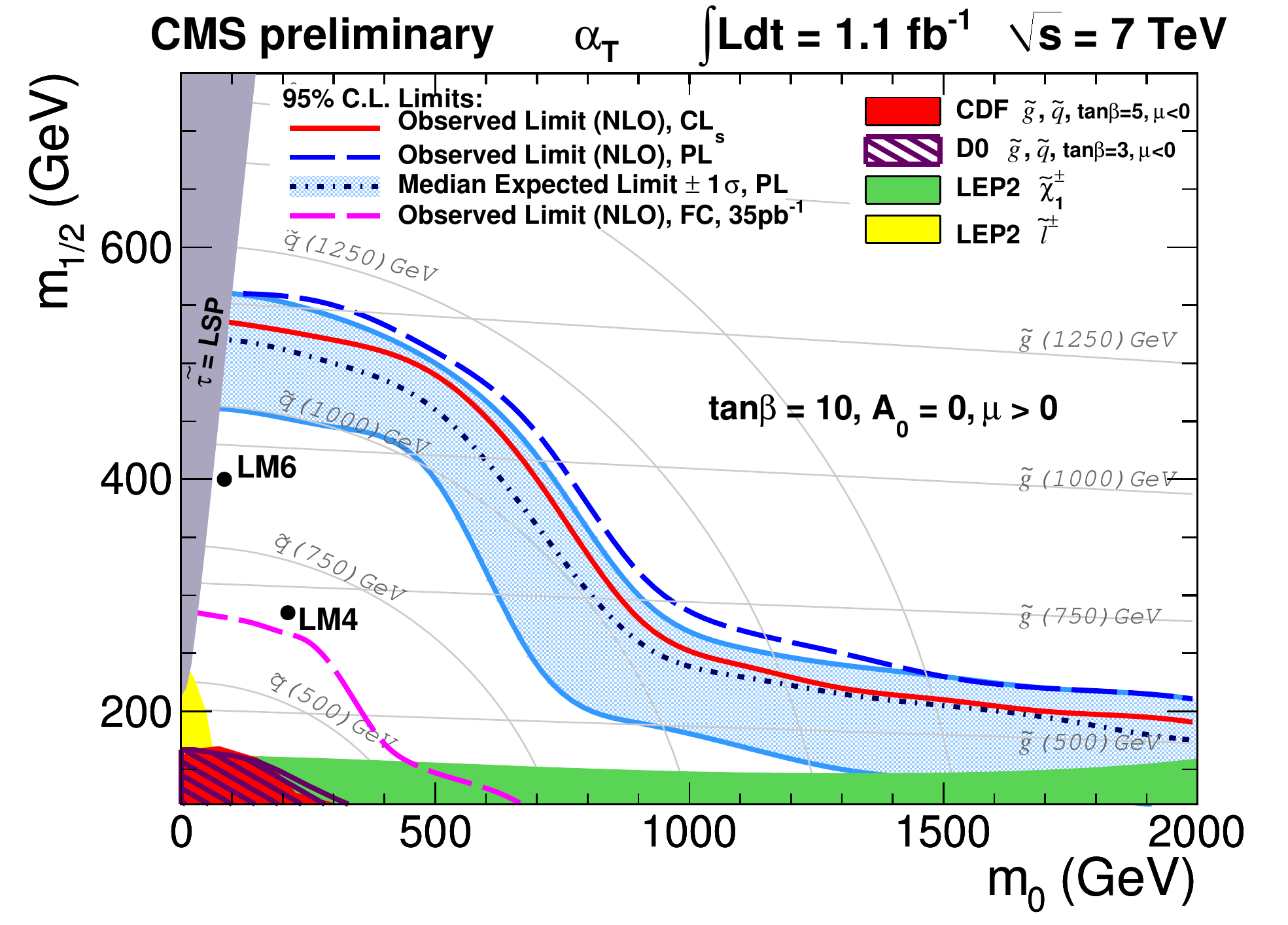}
  \end{minipage}
  \caption{CMS search for SUSY in hadronic final states using
    $\alpha_T$. Left: Comparison of $\alpha_T$ distribution with data
    and expectation. The strong rejection power of $\alpha_T$ against
    QCD multi-jet backgrounds is apparent from the distribution,
    before the $\alpha_T>0.55$ requirement is applied.  Right: The
    null result is interpreted in terms of CMSSM SUSY model. Squark
    and gluino masses of $1.1~\tevcc$ are excluded for $m_0 <
    0.5~\tevcc$.}
  \label{fig:cms_alphat}
\end{figure}
Figure~\ref{fig:cms_alphat} shows the result of a CMS search targeting
a similar final state~\cite{CMS-PAS-SUS-11-003}. The measurement uses
a discriminating variable called $\alpha_T$ to suppress the multi-jet
background.  The variable is defined as
$
  \alpha_T \equiv {E_T^\mathrm{jet2}}/{M_T^{1,2}},
$
where $E_T^\mathrm{jet2}$ is the transverse energy of the subleading
jet and $M_T^{1,2}$ is the transverse mass using the first two jets, and events
are required to pass $\alpha_T>0.55$.  This variable is very effective
at suppressing QCD multi-jet backgrounds, as can be seen in
Fig.~\ref{fig:cms_alphat} left. Figure~\ref{fig:cms_alphat} right
shows the result of this search using 1.1~\invfb of data.  Using an
constrained SUSY model (CMSSM)~\cite{cmssm} to set limits, squark and
gluino masses of $1.1$~TeV are excluded for $m_0 < 0.5$~TeV.
\begin{figure}[tbp]
  \centering
  \includegraphics[width=0.5\textwidth]{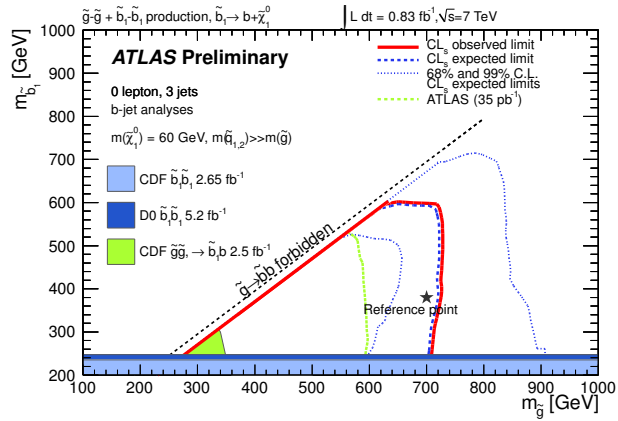}
  \caption{ATLAS results for searches for new physics with $b$ jets +
    \mett using 0.83~\invfb of LHC data. Shown is the excluded region
    for gluino and $b$ squark pair production followed by the decays
    $\tilde{g}\to \tilde{b}_1 b$ and $\tilde{b}_1 \to
    b\tilde{\chi}_1^0$, under the assumption $m_{\tilde{\chi}_1^0} =
    60~\mathrm{GeV}/c^2$. The limits obtained in this measurement
    extend the reach of experiments much beyond the shown Tevatron
    contours.}
  \label{fig:atlas_bjets_met_contour}
\end{figure}

In supersymmetric models, the third family is special.  The top
squarks can be lighter than other squarks due to the large mass of the
top quark.  To a lesser extent, the bottom squark might be much
lighter than the other squarks. Therefore, $\tilde{t}_1$ or
$\tilde{b}_1$ could be produced copiously at the LHC, either directly
or via decays of gluinos The ATLAS collaboration has mounted a search
for $b$ squarks with 0.83~\invfb of data~\cite{ATLAS-CONF-2011-098}.
They consider either the case of direct production or decays through
intermediate gluinos (e.g. $\tilde{g}\to \tilde{b}_1 b$), where
$\tilde{b}_1\to b \tilde{\chi}^0_1$. The analysis requires events with
at least one jet with $p_T>130$~GeV$/c$, an additional jet with
$p_T>50$~GeV$/c$ and $\mett>130$~GeV. The displacement of the common
vertex of tracks associated with a jet with respect to the interaction
point is used to identify $b$ jets. Events with identified leptons are
vetoed. There is no evidence for an excess beyond the SM expectations.
Results on are shown in Fig.~\ref{fig:atlas_bjets_met_contour} as a
function of the gluino mass and bottom squark mass, with the
assumption $m_{\tilde{\chi}_1^0}=60~\mathrm{GeV}/c^2$. Gluinos up to a
mass of 600~\gevcc are forbidden up to $\tilde{b}$ masses at the
kinematic limit. As can be seen from
Fig.~\ref{fig:atlas_bjets_met_contour}, the limits extend much beyond
the shown Tevatron limits with the same model assumptions.


\paragraph{Searches for new gauge bosons}
\begin{figure}[tbp]
  \centering
  \includegraphics[height=2.4in]{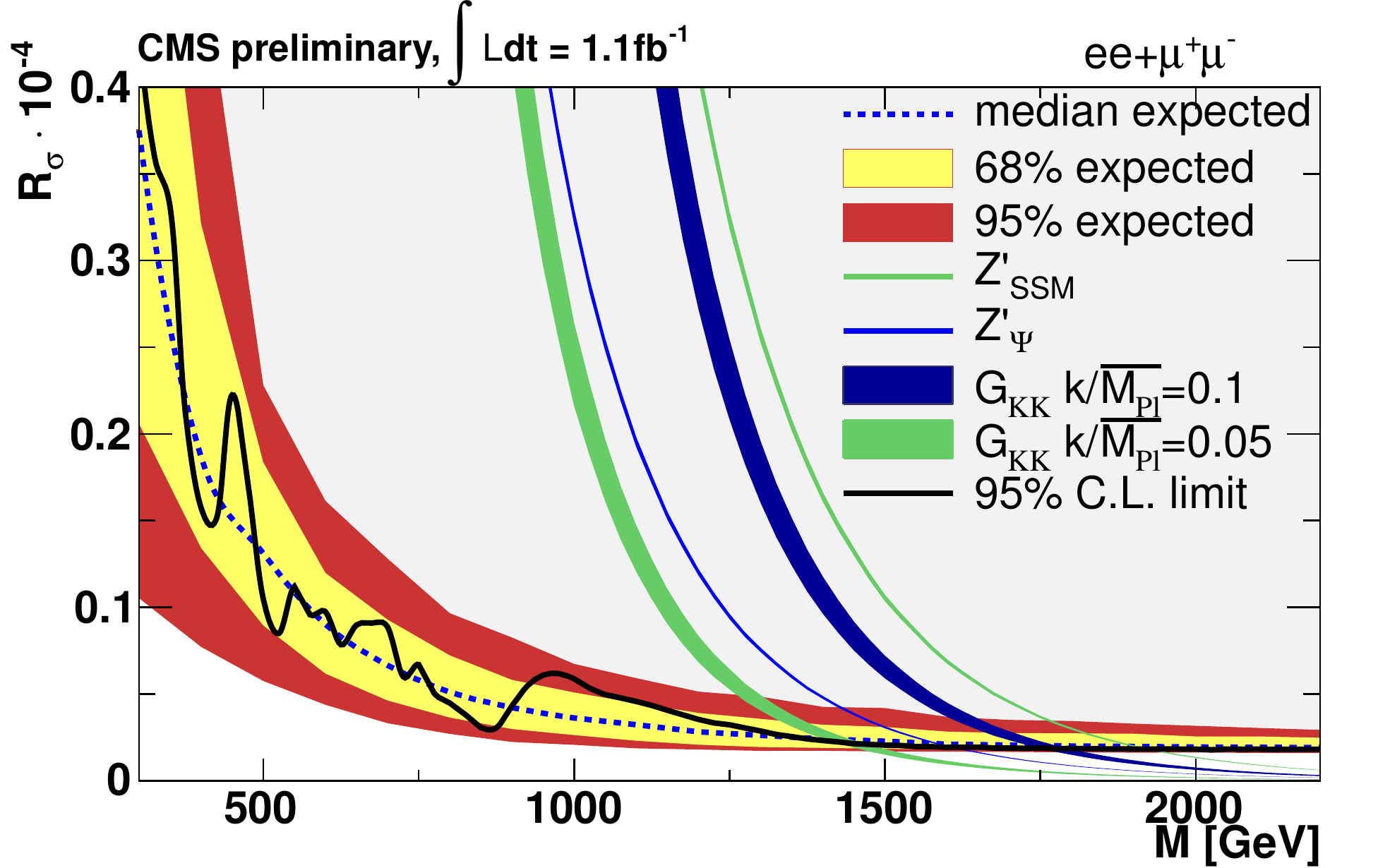}
  \includegraphics[height=2.25in]{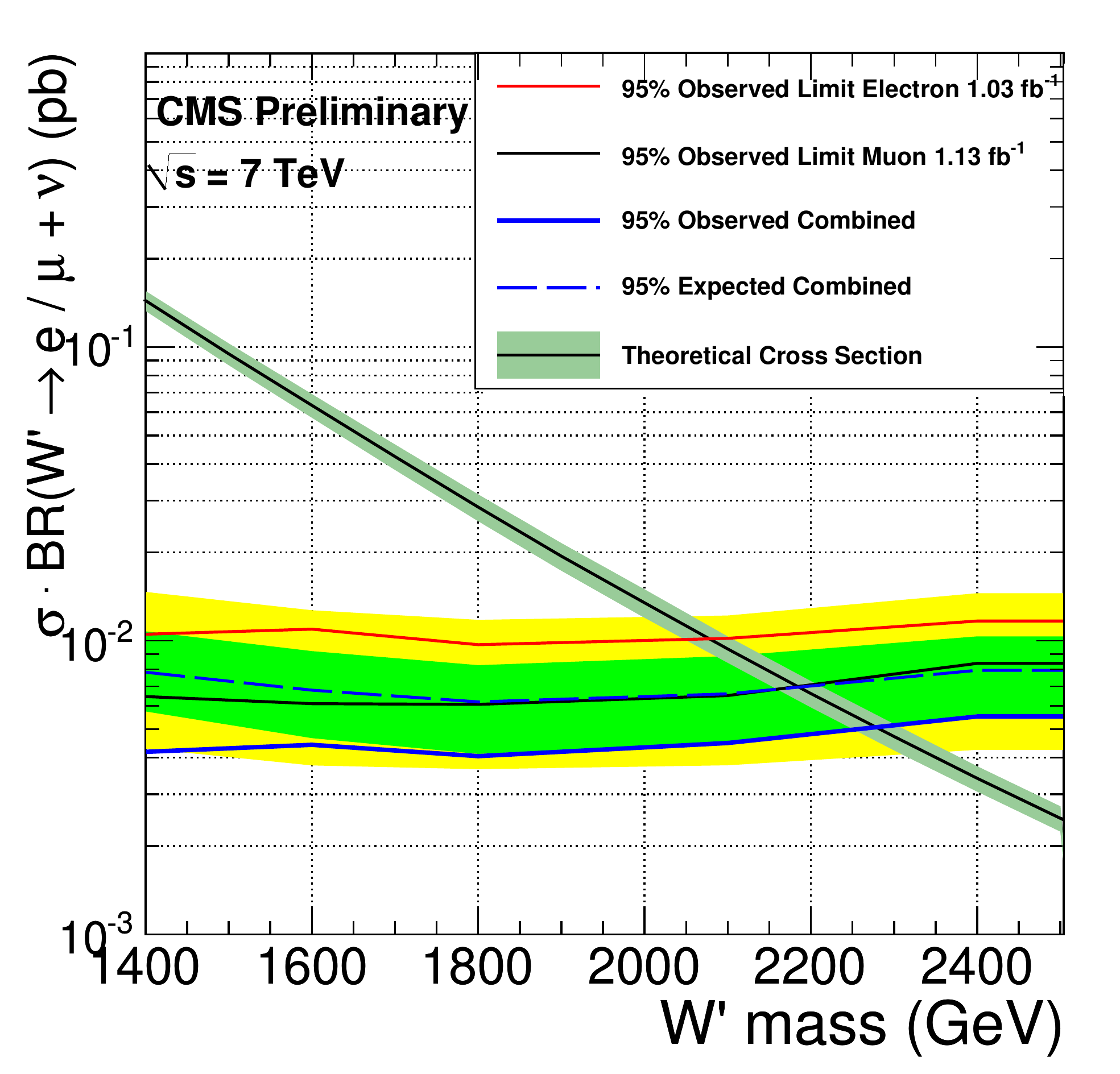}
  \caption{Left: limits on production of narrow resonances that decay
    to di-muon or di-electron pairs, such as $Z^\prime\to \ell\ell$,
    from the CMS experiment. No excess is observed. SM-like heavy new
    $Z$ gauge bosons are excluded for $m<1940~\mathrm{GeV}/c^2$. Right:
    Limits on $\sigma \times BR(W^\prime\to\ell\nu$) from a null
    result from CMS. Limits are placed on SM-like $W^\prime$ for
    $m<2.27~\mathrm{TeV}/c^2$. }
  \label{fig:cms_newgaugebosons}
\end{figure}
Searches for new gauge bosons have been a staple at the Tevatron for
many years. $Z^\prime$ bosons can be found by looking for narrow
dilepton resonances. The CMS collaboration has searched for narrow
di-electron and di-muon resonances with data corresponding to
1.1~\invfb of data. No signal beyond SM background is observed, and
the results are interpreted in terms of SM-like $Z^\prime$ gauge
bosons, Kaluza-Klein gravitons and superstring inspired
models. SM-like $Z^\prime$s are excluded for
$m<1940~\mathrm{GeV}/c^2$, as can be seen in
Fig~\ref{fig:cms_newgaugebosons}~\cite{CMS-PAS-EXO-11-019}.

Searches for heavy copies of the $W$ gauge boson have also broken new
ground. At the time of the conference, the main experimental
$W^\prime$ searches from the LHC were in the leptonic decay mode:
$W^\prime\to\ell\nu$, where $\ell = \mu $ or $e$. Using 1.1~\invfb of
data, the transverse mass distribution is examined in events with a
single isolated high-momentum lepton and large missing energy for an
excess that would signal the decay of a $W^\prime$.  No excess is
found, and heavy boson masses below 2.27 TeV are excluded at the 95\%
confidence level, assuming standard model-like
couplings~\cite{CMS-PAS-EXO-11-024}.
Figure~\ref{fig:cms_newgaugebosons} shows the limits for both $W^\prime$
and $Z^\prime$. Similar results were obtained by the ATLAS
collaboration~\cite{atlaszprime,ATLAS-CONF-2011-109,atlaswprime}.  

\begin{figure}[tbp]
  \centering
  \begin{minipage}{0.5\linewidth}
    \vspace{0pt}
    \includegraphics[width=\textwidth]{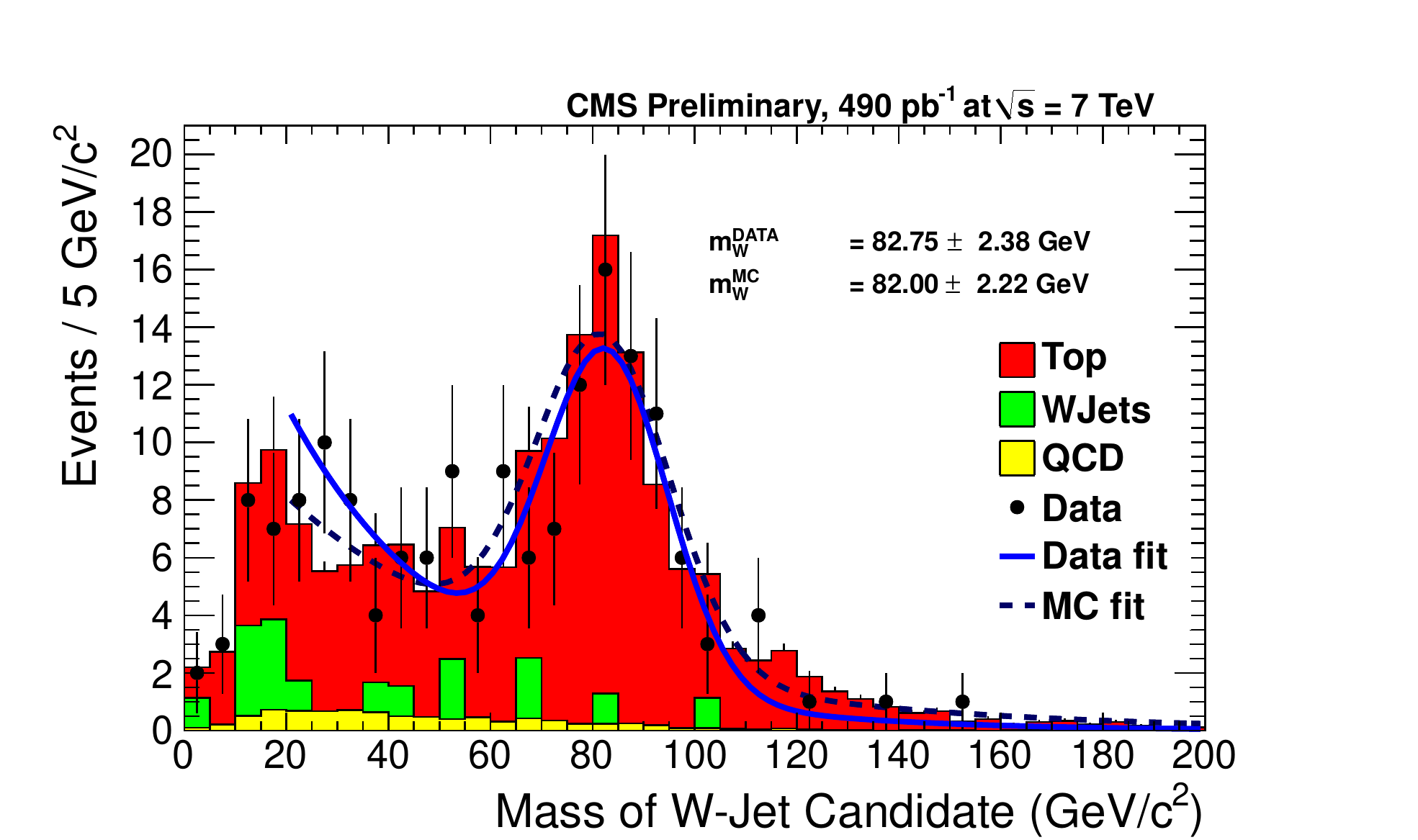}
  \end{minipage}
  \hfill
  \begin{minipage}{0.5\linewidth}
    \vspace{0pt}
    \includegraphics[width=\textwidth]{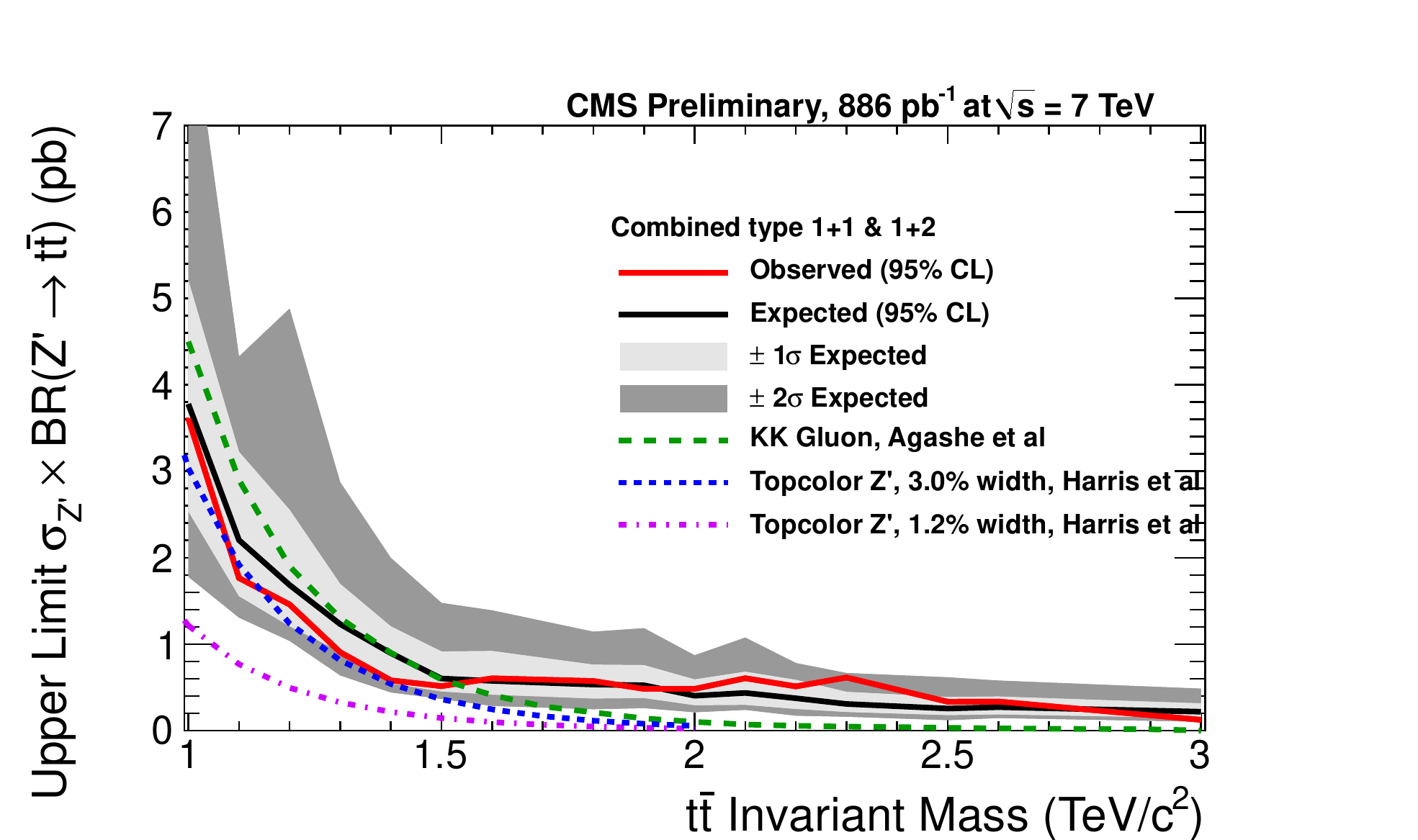}
  \end{minipage}
  \caption{Left: Mass of the highest mass jet in a semileptonic top
    sample, demonstrating the jet substructure technique. Right:
    limits on $\sigma \times \mathcal{B}(Z^\prime\to t\bar t)$, in a
    CMS $Z^\prime$ search.}
  \label{fig:cms_zttbar}
\end{figure}
New gauge bosons can of course also decay in other ways. One search
presented at the conference uses new techniques developed in
collaboration between the theoretical and experimental communities
that focus on the details of the structure of energy deposits
within jets, originally proposed in conjunction with $VH$ searches at
the LHC~\cite{bdrs}.  The paper proposed selecting very boosted Higgs
bosons such that the decay products of the Higgs are reconstructed as
one jet with sub-jets, rather than as two distinct jets, and an
algorithm for detecting such \emph{jet substructure}. Since then many
more techniques based on jet sub- and superstructure have been
proposed, including those to detect boosted top quarks (``top
taggers''), where the $t\to W b$ decay products are not reconstructed
as distinct objects.  The CMS collaboration has performed a search for
$Z^\prime\to t \bar{t}$ process where the top quarks are highly
boosted and therefore the decay products of one or both of the top
quarks is collimated into one jet ($b$ quark, both hadronic $W$ decay
products in one jet) or two jets (both hadronic $W$ decay products in
one jet), rather than the usual three jets~\cite{CMS-PAS-EXO-11-006}.
The use of this technique extends the sensitivity of the search to
higher $Z^\prime$ masses. Figure~\ref{fig:cms_zttbar} left demonstrates
the ability of the top taggers to find $W$ bosons in a top-quark
enriched data sample. No excess is observed beyond SM
expectation. Analyzing 0.865~\invfb of data, CMS sets a limit on
$\sigma_{Z^\prime}\times\mathcal{B}(Z^\prime\to t\bar t)<1$~pb for
$m<1.1$~TeV, as shown in Fig.~\ref{fig:cms_zttbar} right.


\begin{figure}
  \centering
  \captionsetup{type=figure}
  \subfloat[CDF and D0 $W+$~jet events.]{\label{fig:cdf_wjj}
    \includegraphics[width=0.3\textwidth]{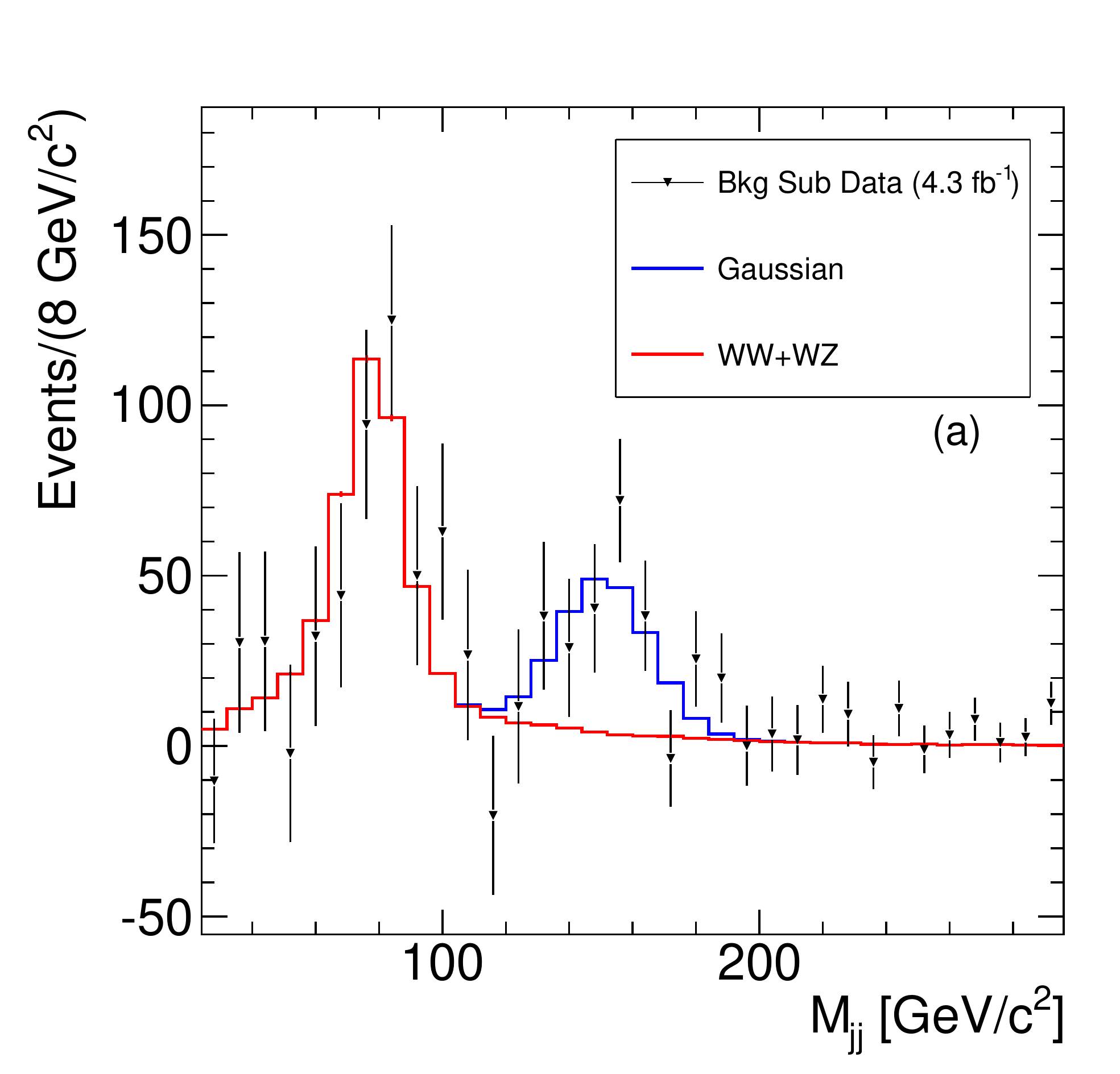}
    \includegraphics[width=0.3\textwidth]{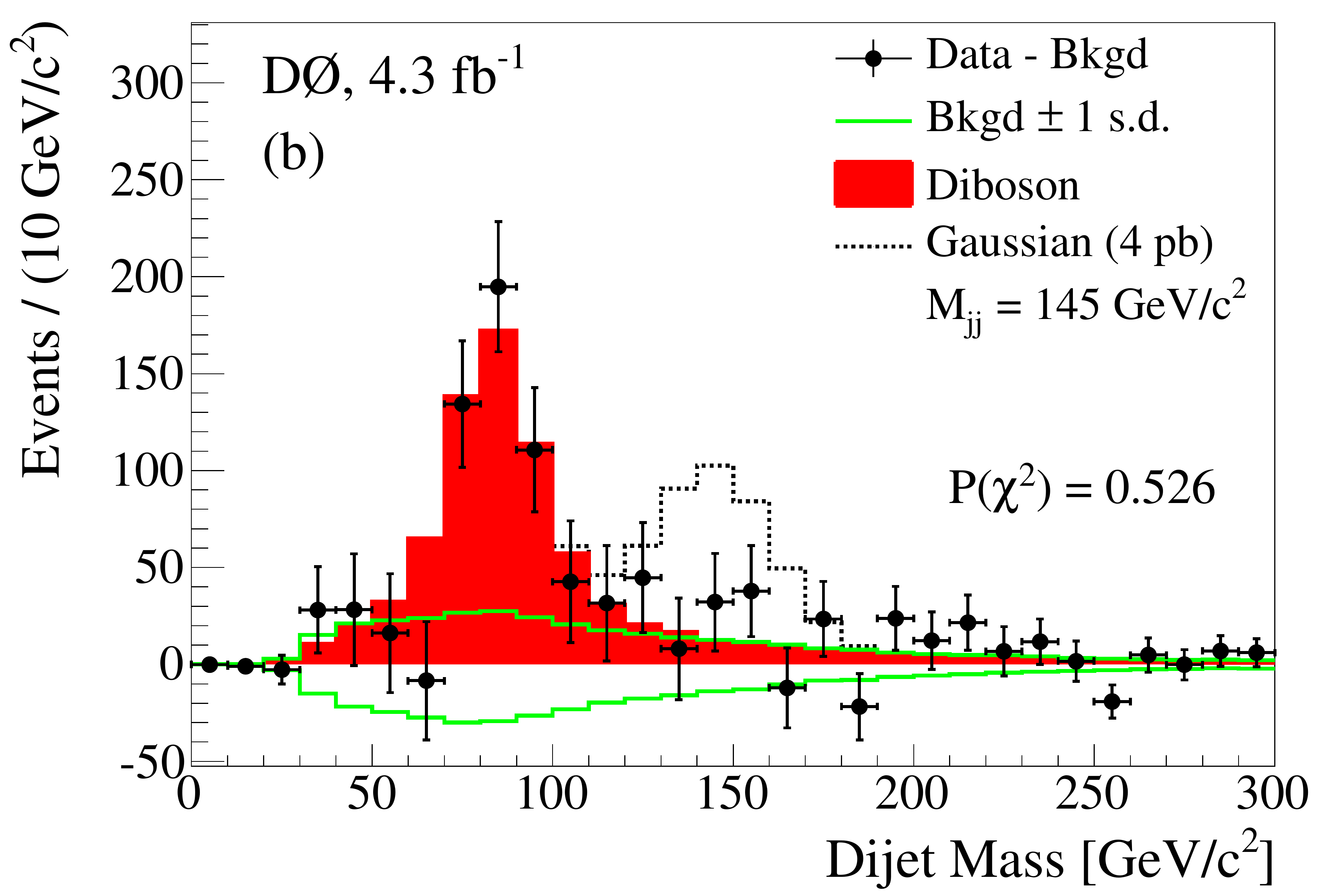}
  }
  \subfloat[D0 Like-sign dimuon asymmetry.]{\label{fig:d0lsmu}
      \includegraphics[width=0.3\textwidth]{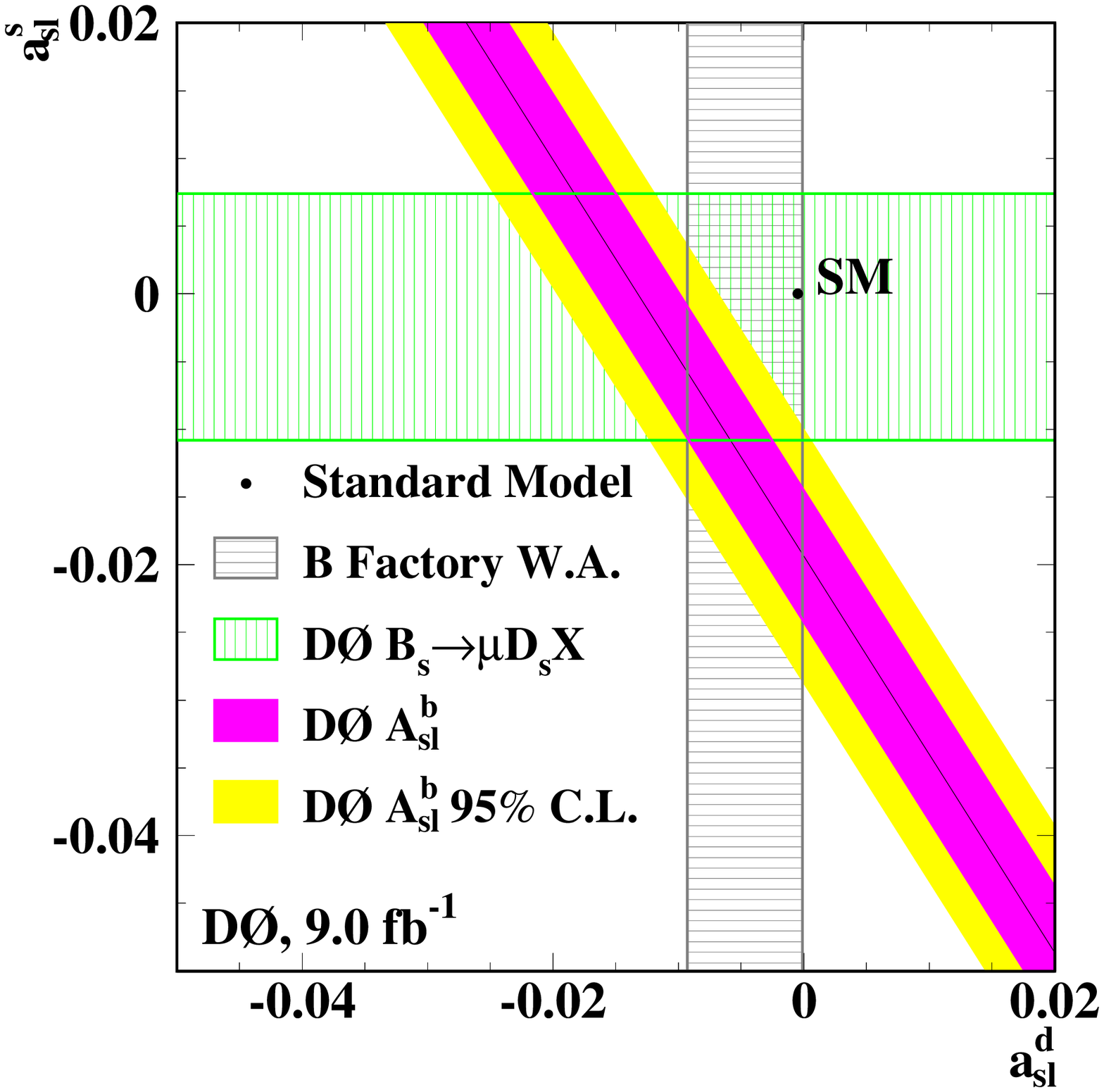}}
    \caption{Left and middle: Background-subtracted invariant mass
      distribution of two jets from a sample of $W+$~jet events. Left:
      CDF. The blue line shows a Gaussian approximation of a new
      physics signal with a resolution compatible with the detector
      resolution and a cross section of approximately three pb. Middle:
      D0. A Gaussian signal with a 4 pb signal is superposed. No
      evidence for a non-standard signal is observed and a CDF-like
      anomaly is ruled out at 2.5 standard deviations.  
      Right: Like-sign dimuon asymmetry measured in semi-leptonic $B$
      meson decays at the D0 experiment in 9~\invfb of data. The
      asymmetry is incompatible with the standard model at $3.9~\sigma$.
    }
\end{figure}

CDF has reported an excess in the invariant mass spectrum of jets in
events with leptonically decaying $W$ bosons~\cite{cdfwjj2011}. In
4.3~\invfb of data, the collaboration reported an excess in the range
$120-160~\mathrm{GeV}/c^2$. The excess was deemed incompatible with
the standard model expectations after taking all known statistical and
systematic effects into account. In a conference result, CDF announced
a result with an increased dataset (7.3~\invfb)~\cite{cdfwjj_7fb} the
significance of the excess increased from 3.2 to 4.1 standard
deviations. Figure~\subref{fig:cdf_wjj} shows the excess from the
publication, as well the result from D0~\cite{d0wjj2011}. Analyzing
the same amount of data as the CDF publication (4.3~\invfb) the D0
analysis checks the CDF measurement and rules out a CDF-like excess at
more than three standard deviations. At the LHC, ATLAS has also
weighed in and cannot confirm the CDF result
either~\cite{ATLAS-CONF-2011-097}.  At this point it is unclear what
the origin of the difference between the CDF and D0 results is, though
the collaborations have released a statement that the data of the two
experiments is compatible with each other within 2.5 standard
deviations.

\paragraph{$A_{fb}$ in $t\bar{t}$ production} CDF and D0 have measured
the forward-backward asymmetry in top quark pair production. In the
Tevatron, the anti-top is produced mostly along the direction of the
anti-proton beam with a small asymmetry predicted by the SM. Both CDF
and D0 have tried to measure this asymmetry in the past and seen a
high value.  CDF reports that the effect is seen to increase as a
function of the invariant mass of the $t\bar{t}$ system and the
rapidity difference~\cite{CDFCollaboration:2010if}.  For
$m_{t\bar{t}}>450~\mathrm{GeV}/c^2$, the parton-level asymmetry is
measured to be $A^{t\bar{t}} = 0.475\pm 0.114$ compared to a
next-to-leading order QCD prediction of $0.088\pm 0.013$.  D0 has
measured a similar though less significant excess in 5.4~\invfb of
data~\cite{Abazov:2011rq}.

At a $pp$ machine like the LHC, there is no intrinsic asymmetry in the
standard model. The forward-backward asymmetry manifests itself as a
discrepancy in the number of forward vs central top quark
pairs~\cite{Hewett:2011vt}. However, estimates suggest that 60~\invfb
are required to allow the LHC to test this. 

\paragraph{Anomalous like-sign dimuon charge asymmetry} D0 has
reported an excess in the number of same-sign muon pairs in
semihadronic $B$ decays using 9~\invfb of data~\cite{d0lsmuon}.
D0 counts the difference between the number of events with two
same-sign positive and negative muons, 
$  A^b_{\mathrm{sl}} \equiv (N^{++}_{b} - N^{--}_{b})/(N^{++}_{b} + N^{--}_{b}),
$
which can be related to the quark-level asymmetries \asld and \asls.
They find an excess compared to the SM prediction:
\begin{eqnarray*}
  A^b_{sl} &=& [-0.787 \pm 0.172 (\text{stat}) \pm 0.093 (\text{syst})]\%\\
  A^b_{sl}(\text{SM}) &=& [ -0.028\,^{+0.005}_{-0.006}]\%
\end{eqnarray*}
The discrepancy is about $3.9~\sigma$. Figure~\subref{fig:d0lsmu} shows
this result compared to SM predictions in terms of the quark-level
asymmetries.

\section{Conclusion}
The PANIC2011 conference took place during an exciting time for
particle physics. Tevatron results with 9~\invfb of data were
presented, showing some interesting hints that the LHC will be able to
follow up on. LHC results with thirty times more statistics than
previously available were also shown. Since then, the LHC has
collected another factor of five times more data.  While no evidence
for new physics has yet to be discovered, it is likely that the next
year will be very interesting.
As mentioned in the introduction, this brief report only allows an
overview of a small number of the results available from ATLAS, CDF,
CMS and D0.  Many more results from all experiments listed are
available on the web~\cite{cms_public_webpage,
  atlas_public_results,cdf_public_results,d0_public_results}.


\begin{theacknowledgments}
  We would like to thank the collaborations for their help and the
  accelerator divisons of CERN and FNAL for their great work. 
  Copyright CERN on behalf of the ATLAS and CMS Collaborations.
\end{theacknowledgments}

\bibliographystyle{aipproc}   

\bibliography{P11_Wittich}

\end{document}